\documentclass[%
 reprint,
 amsmath,amssymb,
 aps,
prb,
floatfix,citeautoscript,longbibliography
]{revtex4-1}

\usepackage{graphicx}% Include figure files
\usepackage{dcolumn}% Align table columns on decimal point
\usepackage{bm}% bold math
\begin{document}

\preprint{APS/123-QED}

\title{Tuning  the Electronic Levels of NiO with Alkali Halides\\Surface Modifiers for Perovskite Solar Cells}

\author{Sofia Apergi$^{1,2}$, Geert Brocks$^{1,2,3}$, and Shuxia Tao$^{1,2}$}
 \email{s.x.tao@tue.nl}
 \affiliation{$^1$Materials Simulation and Modelling, Department of Applied Physics, Eindhoven University of Technology, 5600MB Eindhoven, The Netherlands\\$^2$Center for Computational Energy Research, Department of Applied Physics, Eindhoven University of Technology, P.O. Box 513, 5600 MB Eindhoven, The Netherlands\\$^2$Computational Materials Science, Faculty of Science and Technology and MESA+ Institute for Nanotechnology, University of Twente, P.O. Box 217, 7500 AE Enschede, The Netherlands\\}

\date{\today}

\begin{abstract}
Favorable optoelectronic properties and ease of fabrication make NiO a promising hole transport layer for perovskite solar cells. To achieve maximum efficiency, the electronic levels of NiO need to be optimally aligned with those of the perovskite absorber. Applying surface modifiers by adsorbing species on the NiO surface, is one of the most widespread strategies to tune its energy levels. Alkali halides are simple inorganic surface modifiers that have been extensively used in organic optoelectronics, however, rarely studied in perovskite solar cells. Using density functional theory (DFT) calculations, we investigate the effect of single layer adsorption of twenty different alkali halides on the electronic levels of NiO. Our results show that alkali halides can shift the position of the valence band maximum (VBM) of NiO to a surprisingly large extend in both directions, from $-3.10$ eV to $+1.59$ eV. We interpret the direction and magnitude of the shift in terms of the surface dipoles, formed by the adsorbed cations and anions, where the magnitude of the VBM shift is a monotonic function of the surface coverage. Our results indicate that with alkali halide surface modifiers, the electronic levels of NiO can be tuned robustly and potentially match those of many perovskite compositions in perovskite solar cells.
\end{abstract}

\maketitle

%\tableofcontents

\section{Introduction}
In the last decade, organic-inorganic halide perovskites have emerged as a promising, low-cost alternative to silicon for solar cell applications\cite{Snaith2013}. Their ease of fabrication combined with outstanding optoelectronic properties, such as a tunable optical bandgap\cite{park2013organometal} and a high absorption coefficient\cite{lee2012efficient}, make perovskite solar cells (PSCs) one of the most attractive photovoltaic technologies. Owing to these favorable properties and to the intensive efforts from the scientific community, the power conversion efficiency (PCE) of PSCs has risen rapidly from $3.8 \%$\cite{Kojima2009} to $25.2 \%$\cite{NationalRenewableEnergyLaboratory2019}  over the past ten years. However, the commercialization of PSCs is hindered by their poor long-term stability \cite{Conings2015,Correa-Baena2017,Park2019}.

PSCs comprise a perovskite absorber and several surrounding layers, including electrodes and charge transport layers. The appropriate energy alignment of all layers is important to achieve high PCE, while the stability of each layer and their interfaces determine the overall stability of the device. To find the ideal charge transport layers that meet both the electronic as well as the stability requirements is challenging. For example, the most widely used hole-transport layers (HTLs), such as Spiro-OMeTAD and PEDOT:PSS, yield high PCEs, however lead to long-term instability issues due to their organic nature\cite{You2016}. In order to overcome this issue, attention has been recently drawn to substituting the organic HTLs with affordable inorganic conductors\cite{Li2015}. Because a HTL typically requires a material with a very high work function, many of the common conductors in this role are p-doped metal oxide semiconductors. One such candidate is nickel oxide (NiO)\cite{Subbiah2014}. NiO is a p-type semiconductor with a wide bandgap ($> 3.5$ eV)\cite{Sawatzky1984} and its electronic levels favorably align with the most studied perovskite, MAPbI$_3$\cite{Jeng2014}. Indeed, PSCs using NiO as HTL exhibit an improved photo- and thermal stability\cite{Xu2019} and a PCE that can exceed $20 \%$\cite{Chen2019,Chen2018}.

Despite the advantages of NiO, several challenges remain, including a low hole conductivity\cite{Corani2016}, poor contact\cite{Bai2016} and an unoptimized band alignment between NiO and the perovskite absorber\cite{yin2019nickel}.  A common strategy to overcome the contact problem is by inserting a thin contact layer of a material that improves the chemical bonding between HTL and perovskite\cite{Bai2016}. At the same time, this contact layer modifies the band offset between HTL and perovskite\cite{Goh2007}, which is an important design parameter. The need to optimize this parameter has increased as novel perovskite materials are constantly being developed.

Since the first PSC were made in 2009\cite{Kojima2009}  using MAPbI$_3$ as the absorber, the perovskite formula has now become quite complex. New compositions, like mixed-halide\cite{zarick2018mixed,JIA20191532}, double\cite{jeon2015compositional,Li2016} and triple-cation\cite{Saliba2016,Zhou2019} or 2D perovskites\cite{tsai2016high,Shi2018,Leveillee2018}, have emerged in an attempt to achieve higher efficiency and long-term stability. Aside from that, there are applications, where perovskites with rather different optoelectronic properties are needed. For instance, tandem solar cells, i.e., stacks of Si and perovskite subcells\cite{mcmeekin2016mixed,Bush2017},  or stacks of all-perovskite subcells\cite{Leijtens2018,Zhao2017,Wang2020}, require both wide- and low-bandgap perovskites. This diversity necessitates the ability to properly control the electronic levels and band positions of the HTL for each application\cite{Tao2019}.

Most of the contact layers that have been applied to NiO/perovskite consist of organic molecules, such as DEA\cite{Bai2016} or PTAA\cite{Du2018}, both of which lead to enhanced perovskite crystallinity and an improved electronic level alignment.  Organic layers are likely to lead to similar stability problems as organic HTLs, however. In this respect, inorganic layers may be a more obvious choice. In particular, alkali halides (AXs), may be an interesting option. Alkalis and halogens are already part of PSCs; some of them (Cs, I and Br) as components of the perovskite, while others can be incorporated as dopants (Rb, K, Na, Li, Cl and F), with a positive effect on the stability of the perovskite\cite{Saliba2016a,Dastidar2016,Cao2018,Li2019}. In addition, a variety of AX salts has been extensively used as passivating interlayers in organic optoelectronics\cite{Hung1997,Jabbour1998,Ahlswede2007,Lu2010}. 

In only few studies have AXs been implemented in PSCs, mainly as interface modifiers between the absorber and the TiO\cite{Li2016a} or SnO$_2$\cite{Liu2018} electron-transport layers, while more recently, NaCl and KCl were successfully implemented in PSCs with NiO as the HTL\cite{Chen2019}. In the aforementioned studies, the improved stability and device performance were primarily attributed to defect passivation, whereas changes in the electronic levels of the oxides were observed, but not extensively analyzed. Clearly, NiO surface modification with AXs is a very promising option toward better PSCs and a systematic study is needed in order to understand how these AX layers affect the electronic levels of the oxide.

In this work, we perform density functional theory (DFT) calculations in order to study the effect of AXs on the electronic levels of NiO. Twenty different AXs are adsorbed on the NiO(100) surface and the effect of such an adsorption on the oxide's electronic levels is investigated. The dipoles created at the surface of NiO upon the adsorption of AX result in a shift of the oxide's valence band maximum (VBM). This shift can be either negative or positive, ranging from $-3.10$ eV to $+1.59$ eV. We  show that the direction and magnitude of the shift is determined by the relative position of the adsorbed alkalis and halides, which in turn depends on the specific combination of alkali and halogen and the surface coverage. Our results demonstrate that AXs are excellent surface modifiers for NiO, since they enable us to tune the VBM of the oxide within a wide range, for optimal alignment with any perovskite material for solar cell applications.

\section{Computational Details}

Density Functional Theory calculations were performed using the Projector Augmented Wave (PAW) method as implemented in the Vienna Ab-Initio Simulation Package (VASP)\cite{Kresse1993,Blochl1994,Kresse1996,Kresse1999}. The electronic exchange-correlation interaction was described by the functional of Perdew, Burke, and Ernzerhof (PBE) within the spin-polarized generalized gradient approximation (GGA)\cite{perdew1996generalized}. Energy and force convergence criteria of $10^{-5}$ eV and $2\times10^{-2}$ eV/{\AA} respectively were used in all calculations. In order to properly describe the electronic properties of the strongly correlated NiO, the DFT+U method, as proposed by Dudarev \emph{et al.}\cite{PhysRevB.57.1505,PhysRevB.69.075413} was employed. After extensive testing, a Hubbard parameter of $6.3$ eV and an exchange parameter of $1$ eV was chosen, which corresponds to the values used in previous studies\cite{PhysRevB.57.1505,PhysRevB.69.075413,PhysRevB.74.054421}.

For the geometry optimization of bulk NiO, a unit cell of 4 atoms was used, accounting for the oxide's anti-ferromagnetic (AFM) ordering along the (111) direction. A kinetic energy cutoff of $500$ eV and a $(12\times12\times12)$ ${\Gamma}$-centered $k$-point grid were  employed. The calculated lattice parameter $a = 4.18$ {\AA} is in good agreement with the experimental value, $a = 4.17$ {\AA}\cite{PhysRevB.3.1039}. To model the (100) NiO surface, 4-layer slabs were used with a $2\times2$ square surface supercell, and a vacuum region of $~14\;${\AA} separating the slabs.  The slab calculations were performed with a $(3\times3\times1)$ ${\Gamma}$-centered $k$-point grid and a kinetic energy cutoff of $400$ eV. A  dipole correction was employed to avoid interaction between periodic images\cite{PhysRevB.46.16067}.

The adsorption energies $E_\mathrm{ads}$ of the AXs on NiO were calculated as
\begin{equation}
E_\mathrm{ads} =\left(E_{\mathrm{NiO}/n\mathrm{AX}}-E_\mathrm{NiO}-nE_\mathrm{A}- \frac{n}{2} E_\mathrm{X} \right) / n,
\label{eq:1}
\end{equation}
where $E_{\mathrm{NiO}/n\mathrm{AX}}$, $E_\mathrm{NiO}$, $E_\mathrm{A}$ and $E_\mathrm{X}$ are the DFT total energies of NiO with $n$ adsorbed AX pairs per surface supercell, the clean NiO surface, the alkali metal in its \emph{bcc} structure and the halogen molecules, respectively. Note that, according to Eq.~(\ref{eq:1}), a negative value for $E_\mathrm{ads}$ means that adsorption is energetically advantageous.

For ease of reference, we have also calculated the formation energies of the AX salts 
\begin{equation}
E_\mathrm{AX} = E_\mathrm{AX,total} - E_\mathrm{A} - \frac{1}{2}E_\mathrm{X},
\label{eq:2}
\end{equation}
with $E_\mathrm{AX,total}$ the total energy of the AX salt; \emph{fcc} structures were used for the latter, except for CsCl, CsBr and CsI, which crystallize in a \emph{bcc} structure. For the alkali metals, halogen molecules and AX salts, calculations were performed with the same kinetic energy cutoff and $k$-point grid as in the case of bulk NiO.

\begin{figure}[h]
\includegraphics[width=0.9\linewidth]{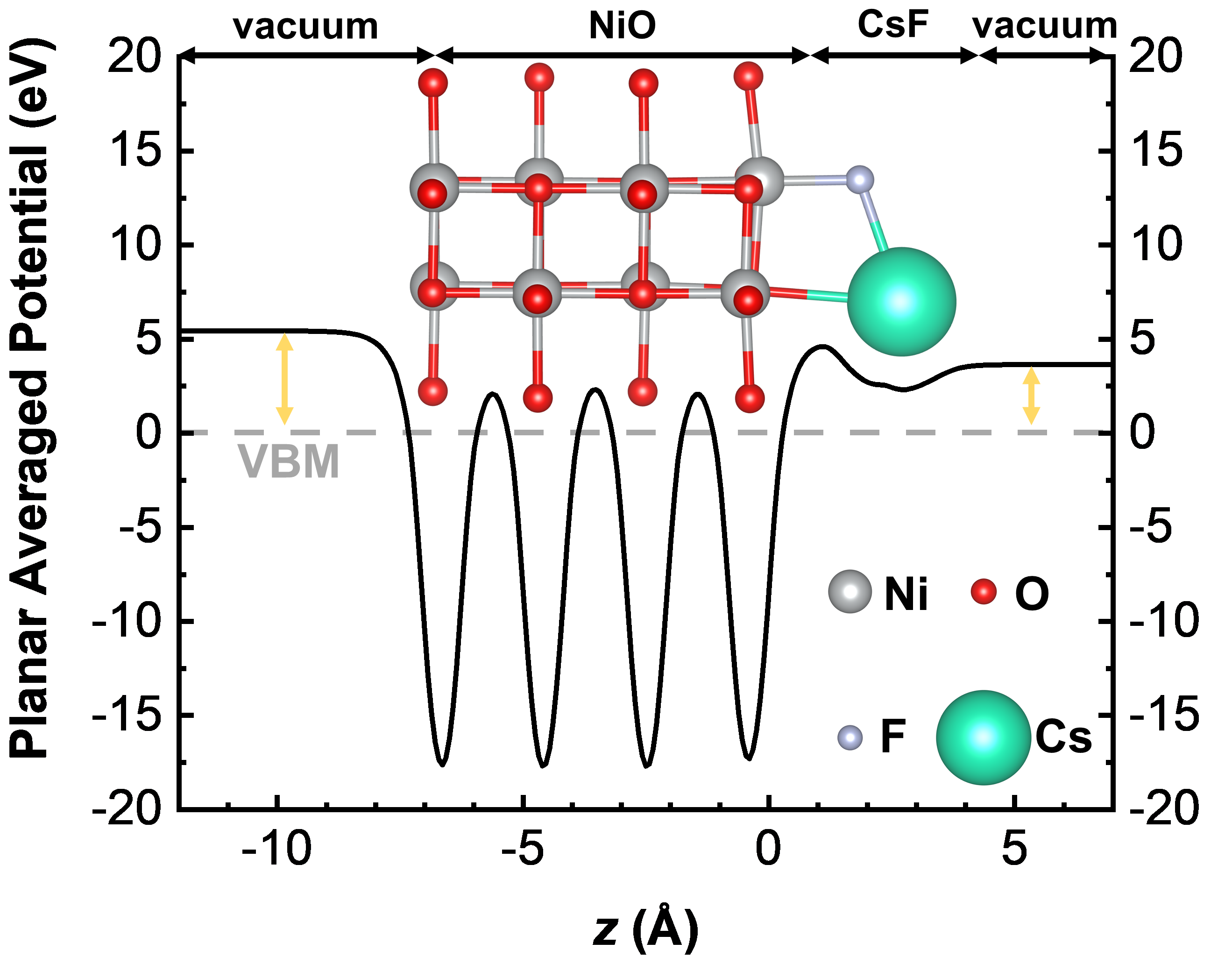}
\caption{\label{fig:1} Planar averaged potential $\bar{V}(z)$, Eq.~(\ref{eq:3}), of a NiO slab covered on one side by 1/8 ML CsF, with respect to the energy of the VBM. The vacuum levels on the clean side of the slab (left), and the CsF covered side (right) are then $5.44$ eV and $3.65$ eV, respectively.}
\end{figure}

The shift of the VBM is defined as the change of the position of NiO's VBM with respect to the vacuum, upon adsorption of AX. Our sign convention is that, when the VBM moves closer to the vacuum level, the corresponding shift is negative. To determine the vacuum level, the electrostatic potential $V(x,y,z)$ was averaged in the $(x,y)$ plane (parallel to the surface of NiO) and plotted as a function of $z$.
\begin{equation}
\bar{V}(z) = \frac{1}{A} \iint_S V(x,y,z)dxdy,
\label{eq:3}
\end{equation}
with $S$ the surface supercell, and $A$ the area of that cell. Approaching the vacuum region, $\bar{V}(z)$ becomes constant, which defines the vacuum level, as illustrated in Fig.~\ref{fig:1}.

\section{Results and Discussion}

The contact layers between the HTLs and the perovskites should be as thin as possible, as they usually consist of insulating materials, and charge transport proceeds via tunneling through these layers. We therefore consider coverages of the NiO(100) surface of $\leq1$ monolayer (ML) alkali halide AX. We study all possible combinations of the alkali and halogen elements, resulting in twenty different AX compounds. To establish possible bonding geometries we first consider a low coverage of 1/8 ML, and then increase the coverage to maximally 1 ML.

\subsection{1/8 ML AX Coverage}

At 1/8 ML coverage we have one alkali and one halogen atom per $2\times2$ NiO(100) surface supercell comprising 8 Ni and 8 O atoms. The most stable positions of the adsorbants correspond to placing the alkali atom over an O atom and the halogen atom over a Ni atom of the substrate. To obtain the most stable bonding configuration, two different cases are tested. In one case, the halogen is placed far from the alkali, so that the interaction between the two is kept to a minimum. In the other case, the distance between the ions is such that a bond can be formed between them. The latter turns out to be the most energetically favorable for all alkali halide combinations. The alkali and halide species adsorb dominantly as ions, with Bader\cite{HENKELMAN2006354} charges of $+0.85\pm0.02 e$ on the alkali ions, and $-0.74\pm0.06 e$ on the halide ions.

\begin{figure}[h]
\includegraphics[width=0.8\linewidth]{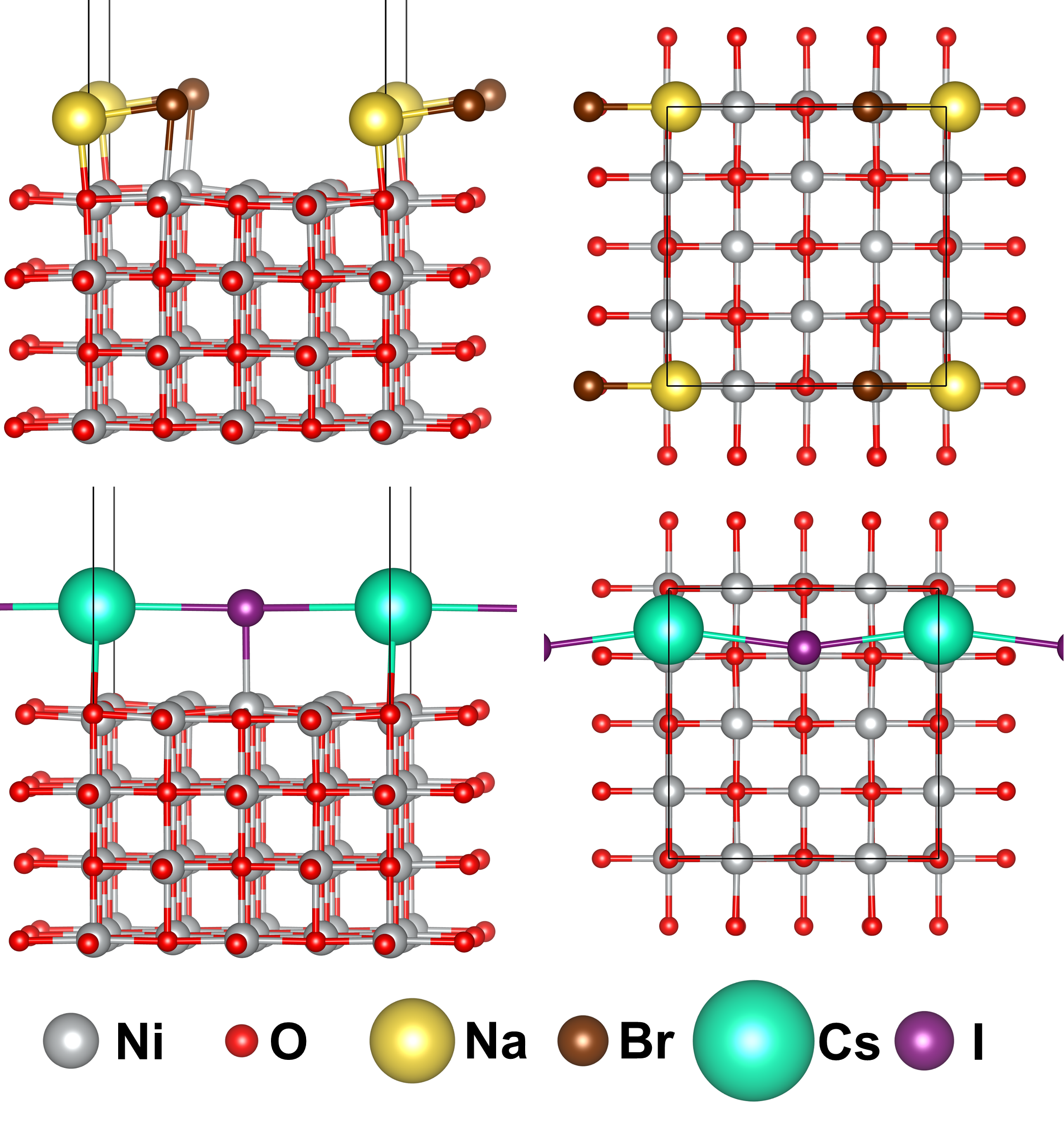}
\caption{\label{fig:2} Geometry of NaBr (top) and CsI (bottom) adsorbed on NiO(100) at 1/8 ML coverage, with side and top views on the left, respectively the right.}
\end{figure}

As an example, Fig~\ref{fig:2} shows the atomic structure of NiO with a NaBr pair adsorbed on its surface. For all twenty AXs, the 1/8 ML adsorption configuration is qualitatively similar to that of NaBr, with the ions adsorbed on nearest neighbor Ni and O atoms, except for CsI, where the large size of the ions requires them to be adsorbed on next-nearest neighbor Ni and O atoms. The nearest neighbor Ni-O distance (2.09 {\AA}) is smaller than the corresponding alkali-halide distance in the AX salts, with the exception of LiF. Not surprisingly then, the bond length of the adsorbed AX pair is $7.7\%$ to $17.2\%$ shorter than in the corresponding AX salt, with the deviations increasing for the larger ions. The exception is again CsI, where the bonding distance between Cs and I is $3.5\%$ larger than in the CsI salt. 

As for the bonds that alkalis and halogens form with O and Ni, they vary from 1.84 {\AA} to 3.23 {\AA} for alkalis and 1.98 {\AA} to 2.87 {\AA} for halogens, increasing with the size of the alkali and halide ions. Depending on the particular A and X combination, the alkali ion A can either be further away from the NiO surface or closer to it, than the halogen ion X. The relative distance of the A and X ions from the surface strongly influences the electronic levels of the oxide, as will be shown below. The adsorption of AXs on NiO has only small effects on the atomic geometry of the oxide's surface, demonstrated by small changes (within 0.21 {\AA})  in the interatomic distances of surface Ni and O atoms. 

\begin{figure}[h]
\includegraphics[width=0.8\linewidth]{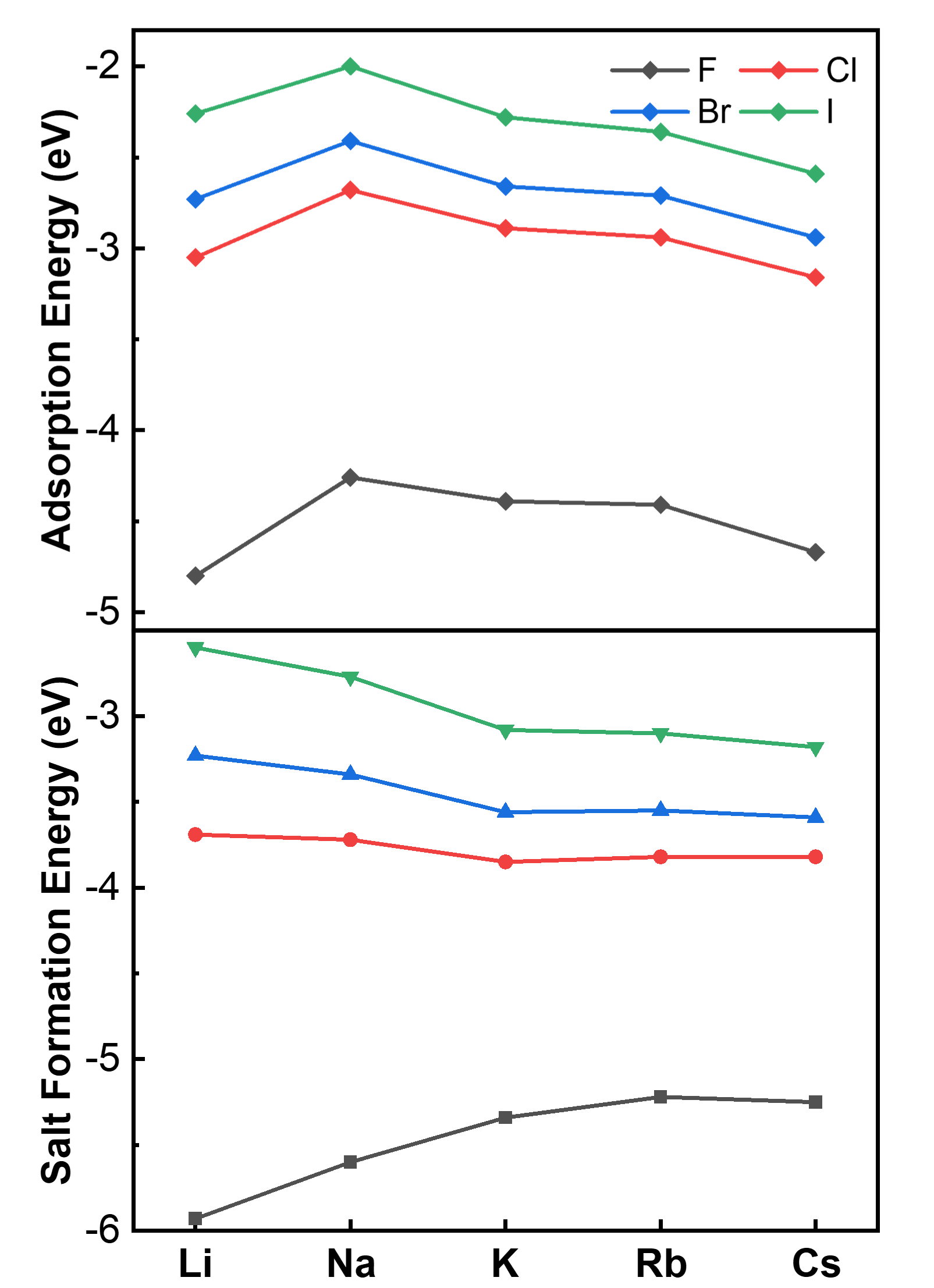}
\caption{\label{fig:3} (a) Adsorption energies $E_\mathrm{ads}$, Eq.~(\ref{eq:1}), (in eV) AX pairs on NiO(100) at 1/8ML coverage vs. A = Li, Na, K, Rb, Cs; the lines guiding the eye are for X = F, Cl, Br, I;  (b) formation energies $E_\mathrm{AX}$ of the AX salts, Eq.~(\ref{eq:2}).  }
\end{figure}

The adsorption energies $E_\mathrm{ads}$, Eq.~(\ref{eq:1}), for the various AX pairs are shown in Fig.~\ref{fig:3}(a). The four halides follow an almost identical trend for varying alkalis.  More specifically, for a given halogen, Na yields the highest adsorption energy, with NaI exhibiting the highest adsorption energy among all AXs ($-2.00$ eV), while the adsorption energy decreases as we move from Na to Cs. Remarkably, the adsorption energies of the fluorides are conspicuously lower than those of the other halides. The adsorption energies for the remaining halides are very close, but for a given alkali, I always yields the highest adsorption energy, followed by Br and then Cl. Of all combinations, LiF yields the lowest adsorption energy of $-4.80$ eV.  

To understand these trends, one can have a look at the formation energies of the AX salts, $E_\mathrm{AX}$, Eq.~(\ref{eq:2}), shown in Fig.~\ref{fig:3}(b). A slight difference between the trends in the AX adsorption energy $E_\mathrm{ads}$ and the AX formation energy $E_\mathrm{AX}$, is that the LiX compounds have a relatively lower $E_\mathrm{ads}$, explained by the strong ionic bond formed between the small Li ion and an O atom of  the substrate. Likewise, in the series from Na to Cs for a fixed halide $E_\mathrm{ads}$ is decreasing (more) than $E_\mathrm{AX}$, which we suggest originates from the alkali becoming increasingly electropositive down the series, making it easier to oxidize it by the substrate, decreasing the energy.

\begin{figure*}[t]
\includegraphics[width=\textwidth]{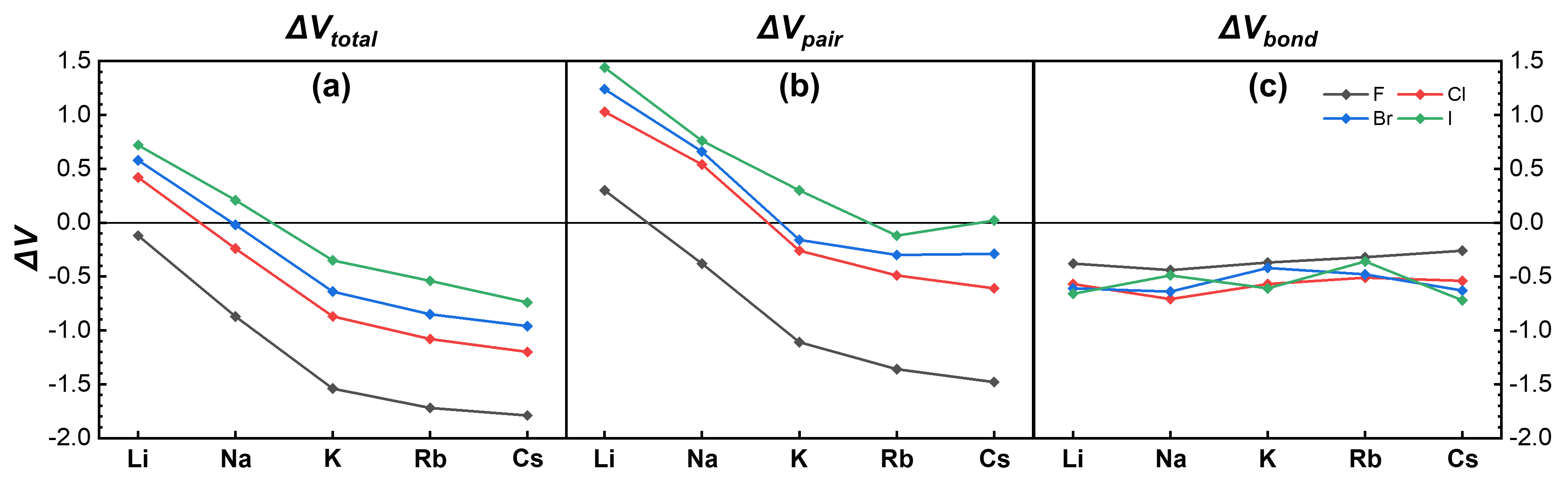}
\caption{\label{fig:4} (a) Total VBM shift ${\Delta}V_\mathrm{total}$, Eq.~(\ref{eq:5}), caused by AX adsorption on NiO(100) at 1/8 ML coverage; (b) Shift ${\Delta}V_\mathrm{pair}$ induced by the AX layer; (c) Shift ${\Delta}V_\mathrm{bond}$ due to charge redistribution upon bond formation between the AX layer and the NiO substrate.}
\end{figure*}

Next, we focus on the effect of AX adsorption on the electronic levels of NiO. The VBM shift  corresponds to the difference in vacuum levels between a clean NiO surface and a NiO surface with an adsorbed AX pair, see Fig.~\ref{fig:1} (in practice, we obtain the two vacuum levels from two separate calculations). Fig.~\ref{fig:4}(a) shows the VBM shift ${\Delta}V_\mathrm{total}$ caused by the adsorption of the various AXs. As can be seen from the graph, the adsorption of different AXs can shift the VBM closer to the vacuum level or further from it, spanning a surprisingly large range of $2.5$ eV. In general, the shift decreases from I to F and from Li to Cs. The shift goes from $+0.72$ eV (LiI) to $-1.79$ eV (CsF), where LiCl, LiBr, LiI and NaI cause positive shifts, NaBr gives an almost zero shift, while the rest of the compounds leads to negative shifts. The VBM shift caused by fluorides stands out by being significantly more negative than that of the rest of the halides. Indeed, fluorides give only negative shifts, whereas other halides can move the VBM to higher or lower energies, depending on which alkali they are paired with. 

In all these  cases the position of the VBM is modified by a potential step resulting from a dipole layer that is created by the adsorption of the AX pairs. It makes sense to split ${\Delta}V_\mathrm{total}$ into two contributions
\begin{equation}
{\Delta}V_\mathrm{total} = {\Delta}V_\mathrm{pair} + {\Delta}V_\mathrm{bond},
\label{eq:4}
\end{equation}
where ${\Delta}V_\mathrm{pair}$ is the potential step resulting from an isolated AX layer of the same density and structure as the adsorbed AX pairs, and ${\Delta}V_\mathrm{bond}$ results from the redistribution of charge that is associated with the bond formation between the AX pair and the NiO surface. The VBM shift ${\Delta}V_\mathrm{pair}$, as calculated with DFT, is shown in Fig.~\ref{fig:4}(b). It is evident that ${\Delta}V_\mathrm{pair}$ follows the same trends as ${\Delta}V_\mathrm{total}$; the shift generally decreases from I to F and from Li to Cs. This means that the overall trend in ${\Delta}V_\mathrm{total}$ is primarily determined by the dipole layer formed by the adsorbed AX pairs. Indeed, ${\Delta}V_\mathrm{bond}$ calculated from Eq.~(\ref{eq:4}), shows a rather uneventful trend around a mean value of $-0.6$ eV.

An independent way to calculate ${\Delta}V_\mathrm{bond}$ starts from the charge displacement originating from the adsorption of the AX pairs. The electron-density difference, ${\Delta}{\rho}$, averaged in the $xy$ plane (parallel to the NiO surface) is given by
\begin{equation}
{\Delta}{\rho}(z) = {\rho}_{\mathrm{NiO}/\mathrm{AX}}(z)-{\rho}_\mathrm{NiO}(z) - {\rho}_\mathrm{AX}(z), 
\label{eq:5}
\end{equation}
where ${\rho}_{\mathrm{NiO}/\mathrm{AX}}$ is the electron density of NiO with AX adsorbed and ${\rho}_\mathrm{NiO}$ and ${\rho}_\mathrm{AX}$ are the densities of the clean NiO slab and the AX pair, respectively, using the same geometry as in the NiO-AX system; ${\Delta}V_\mathrm{bond}$ is then given by
\begin{equation}
{\Delta}V_\mathrm{bond} = -\frac{e}{\epsilon_0 A}\int_{z_0}^{z_1} z \Delta \rho(z) dz,
\label{eq:6}
\end{equation}
where $e$, $\epsilon_0$ and $A$ are the elementary charge, the vacuum permittivity and the surface area of the unit cell, respectively. The lower and upper limit $z_0$ and $z_1$ of the integration are the middle of the NiO slab, respectively the middle of the vacuum region. The values of ${\Delta}V_{bond}$, calculated according to eqs.~\ref{eq:5} and \ref{eq:6}, agree with those calculated with eq.~\ref{eq:4} within $80$ meV.

In all cases, the bond between NiO and the AX causes a negative VBM shift, ranging from $-0.72$ eV to $-0.26$ eV. When AXs are adsorbed on NiO, there is electron accumulation between NiO and the AX and electron depletion in the region of the AX. Such a charge distribution creates a negative potential step and thus shifts the VBM to higher energies. ${\Delta}V_{bond}$ is of similar magnitude for all AXs, with only a slightly lower magnitude for the adsorption of fluorides, possibly because F is much more electronegative compared to the other halogens, which makes extraction of electrons and corresponding depletion at the F site more difficult. 

It should be noted that, in principle, there could be a third contribution to ${\Delta}V_{total}$, Eq.~(\ref{eq:4}), which originates from a change in the geometry of NiO upon the adsorption of AXs. We have calculated this contribution separately and have found it to be minor, ranging from a few meV up to maximally $70$ meV. 

\subsection{Dipole Model}

\begin{figure}[b]
\includegraphics[width=0.8\linewidth]{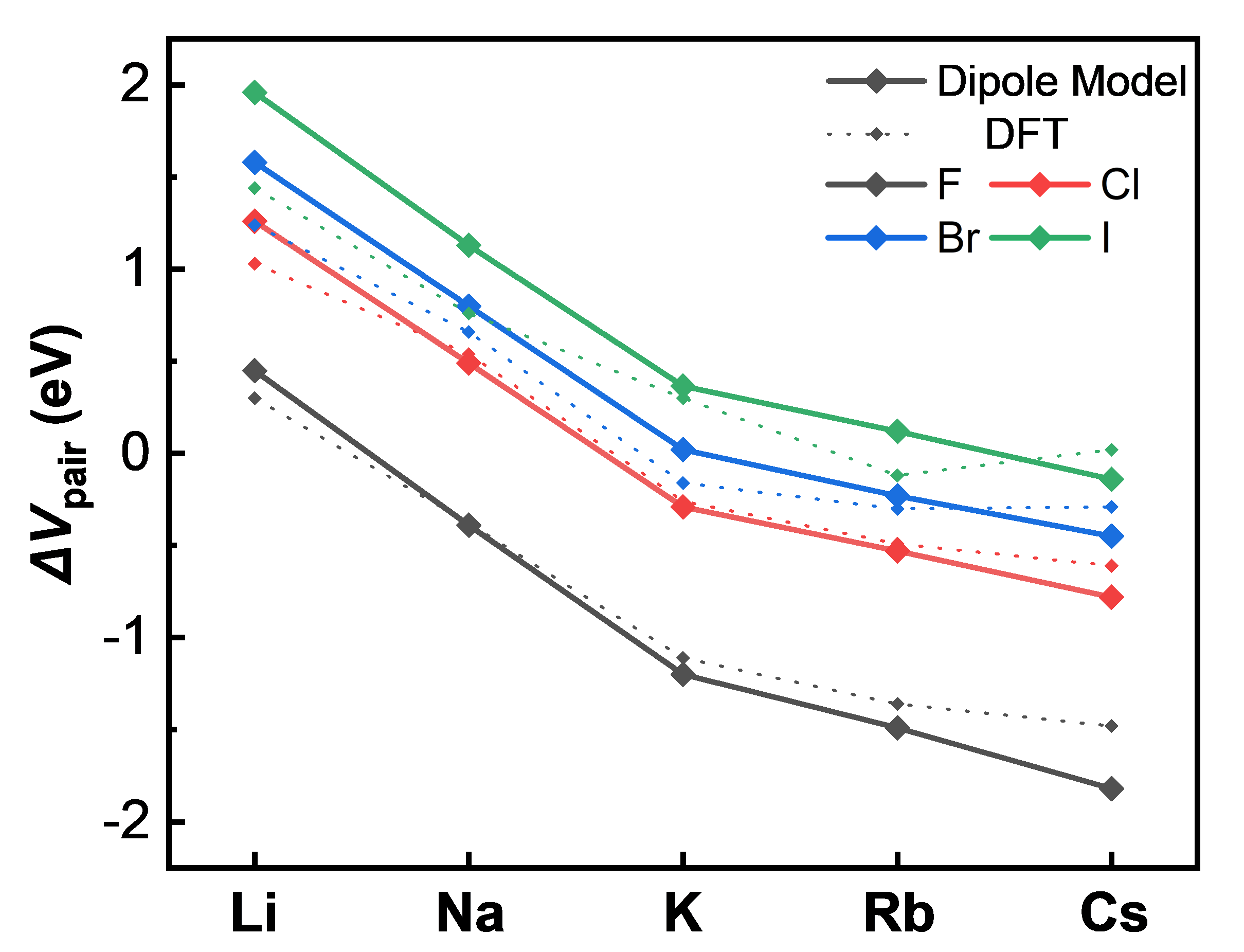}
\caption{\label{fig:5} $\Delta V$ calculated from the point charge model, Eq.~(\ref{eq:7}). The corresponding DFT results are presented for comparison as dotted lines.}
\end{figure}

We conclude from Fig.~\ref{fig:4} that ${\Delta}V_\mathrm{pair}$ dominates the VBM shift ${\Delta}V_\mathrm{total}$. As ${\Delta}V_\mathrm{pair}$ originates from the dipole layer formed by the adsorbed AX species, it may be useful to construct a simple model to explore what distinguishes the various AX species from one another. Starting from a dipole per surface unit cell $\bm{p}$, and selecting the component along the outward normal to the surface, $p_n$ = $\mathbf{p} \cdot \mathbf{\hat{n}}$, then the potential step across the surface (in the outward normal direction) is given by the Helmholtz expression
\begin{equation}
{\Delta}V = -\frac{e p_\mathrm{n}}{\epsilon_0 A}. 
\label{eq:7}
\end{equation}
We model the electric dipole moment $p_\mathrm{n}$ of an AX pair by a pair of opposite charges of magnitude $q$ separated by a distance $d$, $p_\mathrm{n}=qd$, where $d$ is the distance between the ions along the normal direction to the surface of NiO. For $q$ we use the values obtained from the Bader charge analysis. The shifts $\Delta V$, calculated from the simple model of Eq.~(\ref{eq:7}), are shown in Fig.~\ref{fig:5}, and compared with the DFT results for ${\Delta}V_\mathrm{pair}$. 

Clearly, the simple model captures the overall trends in the sign and ordering of the values very well, with only small differences between the exact magnitudes of the shifts. An extended model that takes polarizabilities of the ions into account did not improve the results considerably.  The success of  the simple point charge model proves that the relative position of the A and X ions primarily determines the size of the dipole caused by the adsorbed species, and therefore the shift of the electronic levels of NiO. The trends in $\Delta V$ shown in Fig.~\ref{fig:4} can be interpreted in terms of the distances of the adsorbed anions and cations to the NiO surface, which to a large part are determined by the sizes of these ions. 

If the anion  X is at a distance from the surface that is larger than that of the cation A, the dipole $p_\mathrm{n}$ points toward the surface, and the potential step $\Delta V$ is positive, i.e., it becomes more difficult for electrons to leave the NiO into the vacuum. This situation typically occurs if the anion X is larger than the cation A. The most extreme example of this is LiI, which gives a large $\Delta V \approx 1.5$ eV, see Fig.~\ref{fig:4}. If the anion X is closer to the surface than the cation A, the dipole $p_\mathrm{n}$ points away from the surface. The potential step $\Delta V$ is then negative, and it becomes easier for electrons to leave the NiO into the vacuum. We have this case when the anion X is smaller than the cation A. Here, the most extreme example of this is CsF, which gives $\Delta V \approx -1.5$ eV. If the anion X and the cation A are of similar size, they will be at a similar distance from the surface. The pair dipole is then parallel to the surface, and the normal component $p_\mathrm{n} \approx 0$, which means that $\Delta V \approx 0$. This is the case for KBr and CsI, for instance. 

\subsection{Higher AX Coverage}

\begin{figure}[b]
\includegraphics[width=\linewidth]{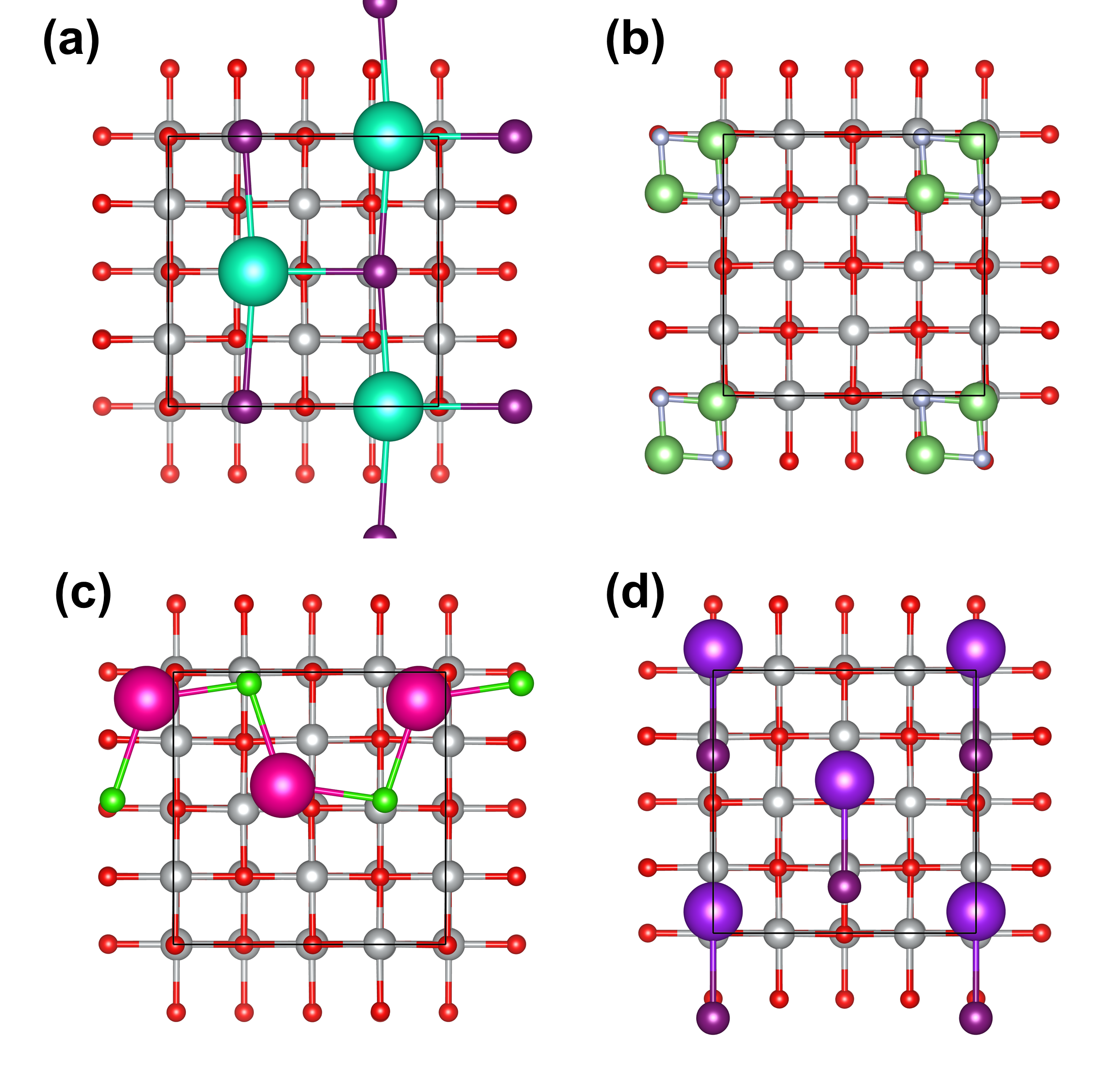}
\caption{\label{fig:6} The four typical adsorption configurations for AX adsorbed on NiO(100) at 1/4ML coverage; (a) CsI; (b) LiF; (c) RbCl; (d) KI}
\end{figure}

\begin{figure}[t]
\includegraphics[width=\linewidth]{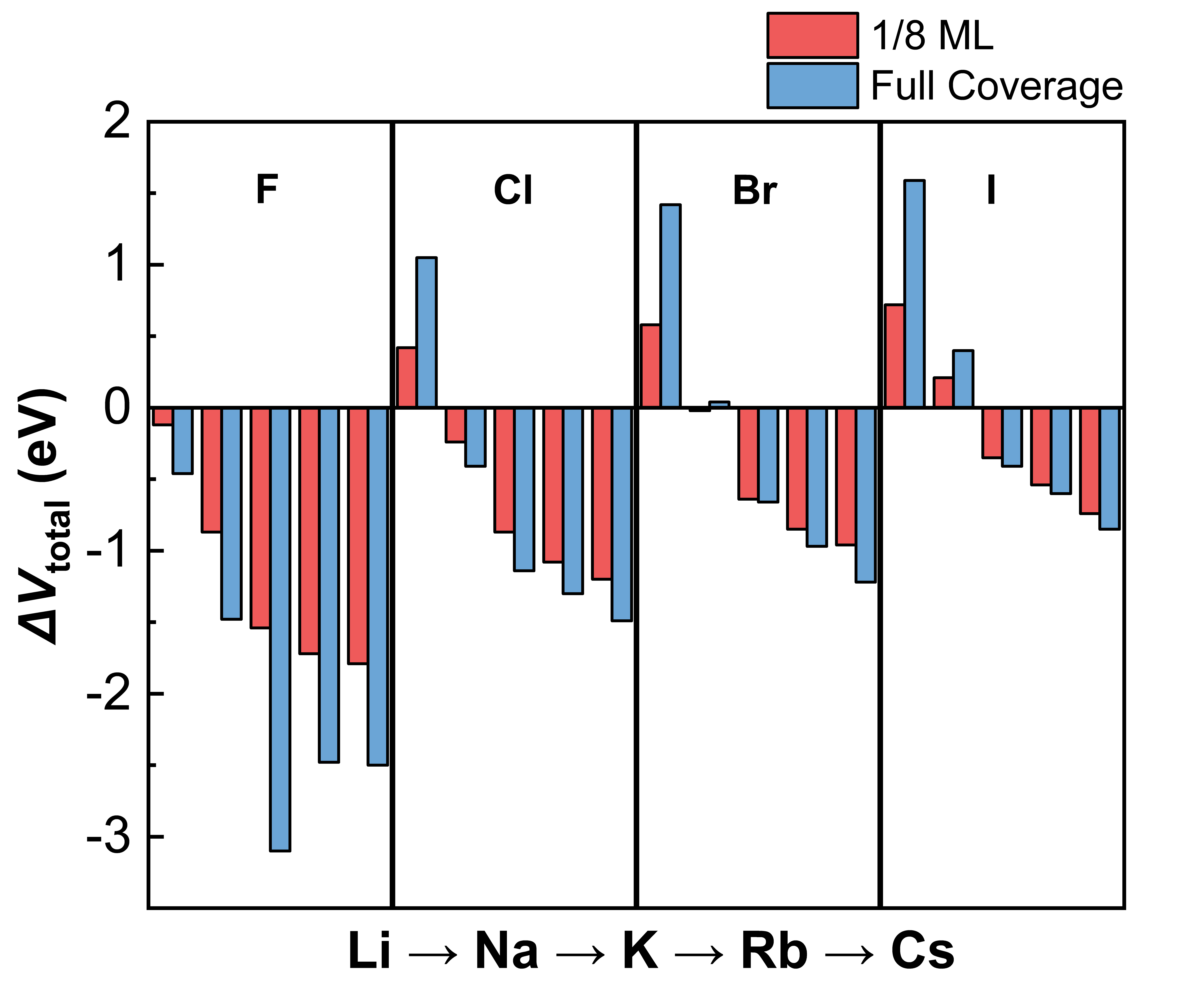}
\caption{\label{fig:7} Total VBM shift ${\Delta}V_\mathrm{total}$, Eq.~(\ref{eq:5}), caused by AX adsorption on NiO(100) at (blue) full coverage (see main text), compared to at 1/8 ML coverage (red).}
\end{figure}

\begin{figure*}[t]
\includegraphics[width=0.8\textwidth]{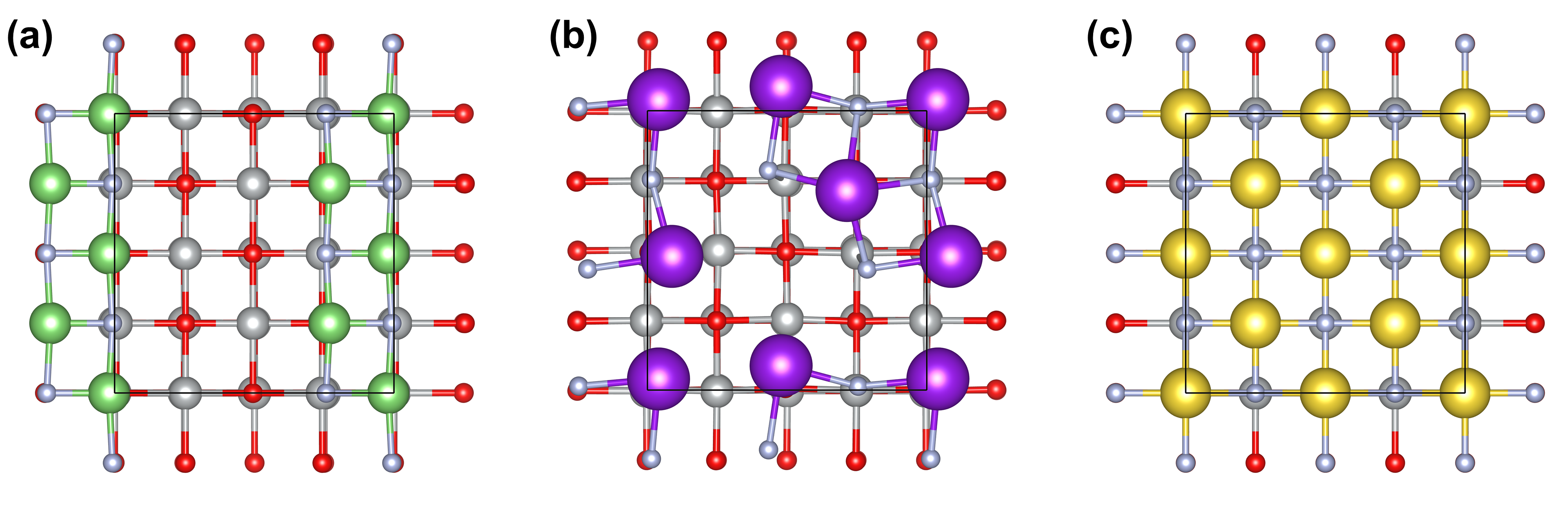}
\caption{\label{fig:8} Adsorption configurations of AX on NiO(100) at (a) 1/2 ML coverage (LiF); (b) 1/2 ML coverage (KF); (c) 1 ML coverage (NaF).}
\end{figure*}

In the previous section we showed that the adsorption of AXs on NiO can shift the VBM of the oxide over a range of $2.5$ eV even at a relatively small coverage of 1/8 ML. Considering the ionic bonding between the A and X cations and anions, and between those ions and the substrate, experimental growth of AX on NiO will likely lead to higher AX densities on the surface. To study the effects of coverage densities, we increase the coverage of the AXs in steps until the maximum that constitutes a closed single layer for each of the AX compounds. Of course, the maximum possible coverage depends upon the size of the A and X ions. For combinations of small A and X ions, such as LiF and NaF, 1 ML coverage can be reached. For the other Li salts, NaCl, and KF, 1/2 ML is the maximum coverage, and for all other AXs, 1/4 ML is the maximum coverage (with the supercell we have used).

For the AX with the largest ions, CsI, doubling the coverage to 1/4 ML already presents a closed AX ML, as illustrated by Fig. \ref{fig:6}(a). For the other AX compounds, four different configurations for 1/4 ML coverage were tested. The most stable configuration depends on the size of the A and X ions and the AX bond length. The smaller A and X ions, which form AX salts with lattice parameters similar to that of NiO, adsorb on neighboring NiO atoms, as shown in Fig. \ref{fig:6}(b). All F, Li and Na salts, as well as KCl, belong in this category. This is not feasible for the remaining ion pairs, because of their larger sizes. In those cases, 1/4 ML coverage results in a frustrated structure, where AX pairs are adsorbed close to neighboring O and Ni atoms, and the pairs link to a zigzag pattern, as shown in Fig. \ref{fig:6}(c). For CsBr and KI, the zigzag pattern involves AX bonds that are parallel to the substrate NiO bond, see Fig. \ref{fig:6}(d). In all adsorption configurations shown in Fig. \ref{fig:6}, the A and X ions cluster, which increases the bonding and leads to a sizable reduction of the adsorption energy, ranging from 0.3 eV to 0.7 eV. Moreover, at 1/4 ML coverage, the AX bond lengths are generally longer than the corresponding bonds at 1/8 ML, but shorter than the bonds in the AX salts, which reflects the transition to a more complete adsorbed monolayer.

Increasing the coverage affects the electronic levels of NiO as well. As demonstrated in Fig.~\ref{fig:7}, the absolute shift of the VBM overall increases for most AXs when the coverage increases. For instance, at 1/4 ML coverage, $\Delta V_\mathrm{total}$ ranges from $-2.50$ eV for CsF to $+0.91$ eV for LiI. Compared with the corresponding numbers for 1/8 ML coverage ($-1.79$ eV; $+0.71$ eV), it is clear that the shifts have become larger. However, the analysis of the contributions to $\Delta V_\mathrm{total}$ and the simple model for the dominant contribution $\Delta V_\mathrm{pair}$, as explained in the previous two sections, still holds at higher coverages. 

Naively, one would expect that  $\Delta V_\mathrm{pair}$ doubles in magnitude when there are twice as many ion-pair dipoles per surface area, according to Eq.~(\ref{eq:7}). In reality, although the contribution is indeed larger for 1/4 ML coverage, it is not twice as large as for 1/8 ML coverage. This is due to the fact that the corrugation of the adsorbed AX layer becomes smaller when going from 1/8 ML to 1/4 ML coverage. This means that the individual AX dipoles $p_\mathrm{n}$ become smaller, Eq.~(\ref{eq:7}). A decreasing corrugation in the AX layer reflects that the interaction within this layer increases with increasing coverage. At the same time, the charge displacement from the NiO substrate to the adsorbed AX also increases somewhat. This is demonstrated by $\Delta V_\mathrm{bond}$, averaged over all AX compounds, goes from $-0.51$ eV to $-0.69$ eV, if increasing the coverage from 1/8 ML to 1/4 ML. 

The trends in the adsorbed geometry of the AX layer going from 1/8 ML to 1/4 ML coverage persist at higher coverages (for the compounds where such a higher coverage in a single layer is possible, of course). At 1/2 ML,the A cations and X anions cluster to form lines, covering one side of the surface of the NiO $2\times2$ supercell, as shown in Fig. \ref{fig:8}(a). Exception is KF, when the adsorbed A and X ions form islands, Fig. \ref{fig:8}(b), which is probably due to K being the largest among the ions that can be adsorbed at this coverage, but it is too large to be embedded in a line structure. Similarly, increasing  the  density  of  the  adsorbed  AX  layer  increases the bonding,  and decreases $E_{ads}$ for most AXs. KF and LiI are exceptions (the adsorption energy is lower for 1/4 ML than it is for 1/2ML for these AXs), but this probably means that the size of these AXs is  near  the  limit  over  which  adsorption  is  not  possible at the  respective  coverage. 

LiF and NaF are the only compounds that can reach 1ML coverage. The Li-F and Na-F bonds are then $2.09$ \AA, the same as the bond lengths between Ni and O atoms of the substrate. Compared to the AX salt, the Li-F bond is then expanded from $2.04$ \AA, while Na-F bond is compressed from $2.35$ \AA. This is accompanied by a further expansion of the bonds between the A and X ions with the substrate, as compared to lower coverages.

Despite these moderate changes in the geometry of the adsorbed AX layer, the shift of the VBM $\Delta V_\mathrm{total}$ generally is a monotonic function of the coverage, i.e., its absolute value increases with increasing coverage. Taking all studied coverages and AXs into account, the VBM can be shifted over a very large range from $-3.10$ eV, with the adsorption of KF, to $+1.59$ eV, for LiI, both at 1/2 ML coverage. 

\subsection{Band alignment to perovskites}

\begin{figure}[h]
\includegraphics[width=\linewidth]{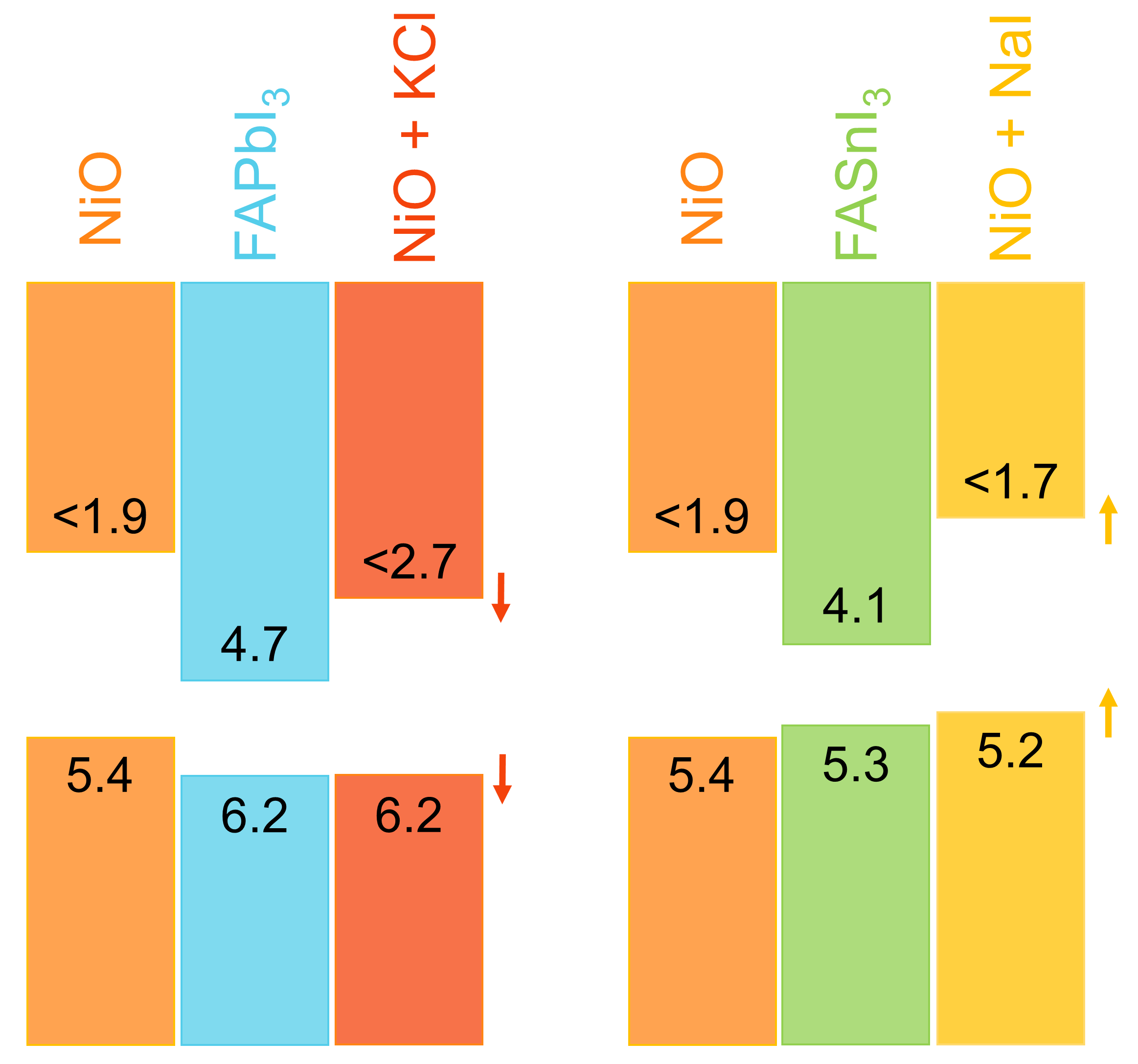}
\caption{\label{fig:9} Band alignment of clean NiO and NiO with AX surface modification with FAPbI$_3$ and FASnI$_3$ perovskites.}
\end{figure}

To put our results into context, we present in Fig. \ref{fig:9} the energy level alignment of NiO with two prototypical perovskites, FAPbI$_3$ and FASnI$_3$, and highlight the effect of alkali halide surface modification. For ease of comparison, we use experimentally determined positions of conduction and valence band maxima and discuss the relative shifts of the levels of NiO upon the adsorption of alkali halides obtained from our DFT calculations. We note here that a wide range of values for the energy levels of NiO and perovskites is reported in the literature due to several factors, such as the fabrication method, the exact composition of the materials, and the data analysis method, determining the exact value of each energy level. Here, we take values from Ref. 13 and Ref. 35 for NiO and the perovskites, respectively. 

As we discussed in the introduction, the valence band maximum (VBM) of NiO aligns reasonably well with that of the most studied perovskite, MAPbI$_3$, the VBM of NiO being 0.5 eV above that of MAPbI$_3$. However, the most efficient and stable solar cells usually are based on different perovskite compositions, such as, FAPbI$_3$ and triple-cation perovskites, FA$_x$MA$_y$Cs$_{1-x-y}$PbI$_z$Br$_{3-z}$. Compared to MAPbI$_3$, the VBM of FAPbI$_3$ lies significantly below that of NiO (0.8 eV), which is far from optimal. To avoid a fundamental voltage loss, the two VBM should be close, $<0.3$ eV, with preferably the VBM of NiO slightly higher in energy. Our calculations indicate many alkali halides (NaCl, NaF, and KX, RbX, CsX, X=I/Br/Cl/F) can potentially induce a downshift of the VBM of NiO, thereby decreasing unfavorably large band offsets and making the extraction of holes more efficient. 

For triple-cation perovskites, FA$_x$MA$_y$Cs$_{1-x-y}$PbI$_z$Br$_{3-z}$, we may assume similar energy level values as in FAPbI$_3$, as the percentages of Cs, MA, and Br in triple-cation perovskites are typically small. This means that alkali halides can also be used to optimize the band alignment of NiO to those perovskites. Ref. 15 presents such an example, where NaCl and KCl were used as interlayers between a triple-cation perovskite and NiO to improve the stability and efficiency of the solar cells. The work concluded the improved performance was coming from the better ordering of the perovskite layer when deposited on top of the NaCl or KCl modified NiO. We suggest additionally that, similar to the case of FAPbI$_3$ (Figure 9), the better energy band alignment due to the presence of KCl could be another reason for the observed improvement in both efficiency and stability, as better charge extraction and transport should mitigate interface degradation and improve device operation stability. 

In contrast to FAPbI$_3$, the VBM of FASnI$_3$ perovskite lies slightly above that of NiO by 0.1 eV. In order for NiO to optimally function as a HTL, the position of its VBM should be ideally pushed slightly upward. For this purpose, NaI could be used, which at 1/8 ML and 1/4 ML (full) coverage moves the VBM of NiO closer to the vacuum level by 0.2 eV and 0.4 eV, respectively. 

We highlight here another very useful perovskite composition, the low band-gap Sn-Pb-based perovskites summarized in Ref. 34, because these perovskites are used in the top cell in all-perovskite tandem cells. The VBM of such mixed compounds retains the character of Sn perovskites \cite{Tao2019}, implying that one can expect similar VBMs. Therefore, the strategy we outlined above for Sn perovskites applies to the Sn-Pb based perovskites as well.

\section{Conclusions}

Nickel oxide is a very attractive hole transport layer for use in solar cells based upon hybrid perovskite  lead halide materials. To minimize the band offset with the range of perovskite materials available, it would be advantageous if one could manipulate the relative position of the valence bands. In this paper we have investigated achieving this goal by adsorption of a layer of alkali halide (AX; A = Li, Na, K, Rb, Cs; X = F, Cl, Br, I) onto NiO, using density functional theory calculations. We find that it is possible to vary the energy position of the NiO valence band maximum (VBM) within a remarkably large range of $4.70$ eV.

The extreme cases are KF, which moves the VBM upwards by 3.10 eV, and LiI, which moves the VBM downwards by 1.59 eV, with all other alkali halides leading to values within this interval. Analyzing the VBM shift, we find that the dominant contribution emerges from the dipole layer that is formed by the adsorbed alkali halide species. Representing this layer with a simple dipole model, we show that the direction and magnitude of the shift mainly originate from the relative distance with respect to the NiO surface of the adsorbed ions, as well as their adsorption density.

Our results indicate that alkali halides are promising surface modifiers for NiO, providing us with the ability to control the valence band offset between NiO and perovskite materials, which in turn could lead to better, more efficient conversion of solar energy.

\begin{acknowledgments}
S. Apergi acknowledges supports from funding NWO START-UP from the Netherlands. S. Tao acknowledges funding by the Computational Sciences for Energy Research (CSER) tenure track program of Shell and NWO (Project Number 15CST04-2) as well as NWO START-UP from the Netherlands.
\end{acknowledgments}

\bibliography{NiO_paper}% Produces the bibliography via BibTeX.

\end{document}